\documentclass[aps,prb,twocolumn,showpacs]{revtex4-1}
\usepackage{latexsym}
\usepackage{amssymb}
\usepackage{graphicx}
\usepackage{amsmath}
\usepackage{bm}
\usepackage{comment}
\begin{document}
\title{Low-temperature linear transport of two-dimensional massive Dirac fermions in silicene: residual conductivity and spin/valley Hall effects}
%\begin{comment}
\author{Yuan Yao}
%\email{andsee134@hotmail.com}
\affiliation{Department of Physics and Astronomy, Shanghai
  Jiaotong University, 800 Dongchuan Road,
  Shanghai 200240, China}
\affiliation{Collaborative Innovation Center of Advanced Microstructures, Nanjing University, Nanjing 210093, China}
\author{S. Y. Liu}
\email{liusy@sjtu.edu.cn}
\affiliation{Department of Physics and Astronomy, Shanghai
  Jiaotong University, 800 Dongchuan Road,
  Shanghai 200240, China}
\affiliation{Collaborative Innovation Center of Advanced Microstructures, Nanjing University, Nanjing 210093, China}
\author{X. L. Lei}
\affiliation{Department of Physics and Astronomy, Shanghai
  Jiaotong University, 800 Dongchuan Road,
  Shanghai 200240, China}
\affiliation{Collaborative Innovation Center of Advanced Microstructures, Nanjing University, Nanjing 210093, China}
%\end{comment}

\begin{abstract}

Considering finite-temperature screened electron-impurity scattering, we present a kinetic equation approach to investigate transport properties of two-dimensional massive fermions in silicene. We find that the longitudinal conductivity is always nonvanishing when chemical potential lies within the energy gap. This residual conductivity arises from interband correlation and strongly depends on strength of electron-impurity scattering. We also clarify that the electron-impurity interaction makes substantial contributions to the spin- and valley-Hall conductivities, which, however, are almost independent of impurity density. The dependencies of longitudinal conductivity as well as of spin- and valley-Hall conductivities on chemical potential, on temperature, and on gap energy are analyzed.

\end{abstract}


\maketitle

\section{ Introduction}

In the last decade, the electronic properties in graphene, a single-atom-thick two-dimensional (2D) layer of carbon atoms in a hexagonal honeycombed lattice, have been extensively studied both theoretically and experimentally.\cite{Novoselov22102004,Novoselov2005,Zhang2005,PhysRevLett.94.176803,RevModPhys.81.109,RevModPhys.83.407,RevModPhys.86.959}
In this system, the low-energy carriers near two nodal ("Dirac") points in the Brillouin zone possess linear energy spectrum and they behave just as massless, two-dimensional, relativistic Dirac fermions.\cite{PhysRev.71.622,PhysRevLett.53.2449,PhysRevLett.61.2015,PhysRevLett.53.2449,PhysRevLett.61.2015} The high mobility as well as a long mean
free path at room temperature, makes graphene a promising
candidate for future electronic applications.

However, the application of graphene in spintronics is quite limited: due to the weak spin-orbit coupling (SOC), the energy gap in graphene is very small.\cite{PhysRevB.75.041401} Recently, graphene's silicon analog, silicene, has been proposed\cite{PhysRevB.76.075131,PhysRevB.79.115409}
and synthesized.\cite{APL.96.261905,APL.97.223109,APL.98.081909,PhysRevLett.108.155501,PhysRevLett.108.245501,1882-0786-5-4-045802,Kara20121} It also has Dirac cones which are similar to those of graphene, but its energy gap due to intrinsic SOC at the Dirac points may
reach the value about $2\Delta_{\rm SO}=1.55\sim 7.9$\,meV ($\Delta_{\rm SO}$ is the characteristic energy of this SOC).\cite{PhysRevLett.107.076802,PhysRevB.84.195430} Besides, buckled lattice structure with two different sublattice planes separated by a vertical distance enables us to break the inversion symmetry in silicene by applying an external perpendicular electric field. This makes another possibility to control the energy gap in silicene (this energy gap is denoted by $2\Delta_z$).\cite{doi:10.1021/nl203065e,PhysRevB.85.075423,1367-2630-14-3-033003,PhysRevB.88.245408} Tunning the values of $\Delta_{\rm SO}$ and $\Delta_z$,  
a phase transition from a quantum spin-Hall state ($|\Delta_{z}|<\Delta_{\rm SO}$) to a trivial insulating state ($|\Delta_{z}|>\Delta_{\rm SO}$) is expected to be observed.\cite{PhysRevLett.107.076802,APL.102.162412,PhysRevB.86.195405,PhysRevB.87.235426,APL.102.043113,PSSR:PSSR201206202,0953-8984-26-34-345303} It was reported that, such a transition from the topological
insulator (TI) to the trivial band insulating (BI) state produces a quenching of the quantum spin Hall effect (SHE) and an onset of an analogous quantum valley Hall effect (VHE).
In Refs.\,\onlinecite{PhysRevB.87.235426} and \onlinecite{0953-8984-26-34-345303}, the intrinsic SHE and VHE induced by ac electric field also have been examined. However, up to now, the effect of electron-impurity interaction on these phenomena has not been analyzed yet. The previous studies on spin-Hall effect in conventional 2D systems with Rashba and/or Dresselhaus spin-orbit couplings\cite{PhysRevLett.83.1834,Murakami05092003,PhysRevLett.92.126603} indicated that the contribution of electron-impurity scattering to spin-Hall conductivity is quite important. It even can lead to the vanishing of total spin-Hall conductivity in conventional 2D electron gas with Rashba SOC.\cite{PhysRevB.70.041303,PhysRevLett.93.226602,PhysRevB.71.245318,PhysRevB.71.033311,PhysRevB.71.245327,PhysRevB.73.035323} Hence, to investigate the SHE and VHE in silicene, extrinsic mechanism associated with electron-impurity interaction is expected to play a substantial role. 

In previous studies on transport in graphene, one of the most intriguing phenomenon is the residual conductivity observed in carrier-density dependence of conductivity: the conductivity reduces to a finite value of order of $e^2/h$ when chemical potential tends to zero.\cite{Novoselov22102004,Novoselov2005,Zhang2005,PhysRevLett.94.176803} Much theoretical effort has been devoted to quantitatively explain this minimum longitudinal conductivity. It is generally accepted\cite{Adam20112007,PhysRevLett.99.176801,PhysRevLett.103.236801,PhysRevB.76.233402,PhysRevLett.101.166803,PhysRevB.79.245423} that the origin of residual conductivity is associated with the formation of electron-hole puddles\cite{Martin2007,Zhang2009} in graphene, induced by randomly distributed charge impurity when the global average carrier density is low. However, in clean samples, such as in suspended graphene samples, where the charged impurities are removed upon annealing and puddle formation should be suppressed, the residual conductivity still can be observed\cite{PhysRevLett.101.096802,Bolotin2008351,Du2008} and hence the mechanisms essentially independent of disorders are required. In previous studies, two such mechanisms were proposed. One is the interband correlation:  notwithstanding the vanishing of equilibrium electron density, dc electric field excites an electron from the valence to conduction band, resulting in a nonvanishing conductivity.\cite{PhysRevB.76.235425,PhysRevLett.99.216602,Kailasvuori2013,PhysRevB.78.235417,JAP.104.043705,Kailasvuori2010,raey,1402-4896-81-3-035702} The other one is associated with the low-energy states at the edges of samples. It was demonstrated that this edge-state mechanism makes sizable contribution to the subgap conductance in bilayer graphene even for highly imperfect edges.\cite{Li2010}
Although the residual conductivity in gapped systems, such as in bilayer graphene,\cite{Li2010,Katsnelson2006,PhysRevB.77.115436,PhysRevB.82.075423,Feldman2009,PhysRevB.81.161407,PhysRevB.82.155308,0295-5075-98-4-47007} has been extensively studied, but it is still unclear in silicene up to now.

In present paper, considering interband coherence, we present a pseudohelicity-basis kinetic equation approach to investigate the effects of electron-impurity interaction on longitudinal conductivity as well as on the spin- and valley-Hall conductivities in both the TI and BI states of silicene. We find that, in addition to the intrinsic SHE and VHE, electron-impurity interaction makes substantial contributions to spin- and valley-Hall conductivities (SHC and VHC) although these contributions are practically independent of impurity density. We also clarify that the longitudinal residual conductivity can reach the value of order of $e^2/h$ at low temperature if the scattering is relatively weak. The dependencies of residual conductivity and of spin- and valley-Hall conductivities on temperature, on band gap and on chemical potential are analyzed by considering finite-temperature screened electron-impurity scattering. 

The paper is organized as follows. In section II, we present the kinetic equation in pseudo-helicity basis. The numerical results are shown in Section III.  Finally, we conclude our results in section IV and append the results within relaxation time approximation in Appendix.   

\section{Theoretical Formulation}

Consider a 2D massive Dirac fermion with momentum ${\bf k}\equiv (k_x,k_y)$ and electric charge $-e$ in buckled silicene. Its motion near the $K$ or $K'$ Dirac node can be described by a Hamiltonian of the form [$\lambda_{\eta\sigma}=\Delta_{z}-\eta\sigma\Delta_{\rm SO}$]
\begin{eqnarray}
\check h_{\eta\sigma} ({\bf k})&=&v_F(\eta k_x\hat\tau_x+k_y\hat\tau_y)+\lambda_{\eta\sigma}\hat\tau_z,\label{H1}
\end{eqnarray}
with $\eta=\pm 1$ as valley index for $K$ and $K'$, and $\sigma=\pm 1$ as spin index for spin up and down. $\hat\tau_i$ ($i=x,y,z$) represent the Pauli matrices, $v_F\approx 0.5\times 10^{6}$\,m/s is the Fermi velocity of Dirac fermion in silicene and the characteristic energy of intrinsic SOC is chosen to be $\Delta_{\rm SO}\approx 3.9$\,meV.\cite{PhysRevB.84.195430,1367-2630-14-3-033003} $\Delta_z$ comes from a hybridization of $p_z$ orbitals with $\sigma$ orbitals of silicon atoms and its value can be tunned by applying external electric field along $z$-direction.\cite{doi:10.1021/nl203065e} Hamiltonian (\ref{H1}) can be diagonalized analytically: the eigen wavefunction $\Psi_{\eta\sigma\mu{\bf k}}({\bf r})$ ($\mu=\pm$) can be written as $\Psi_{\eta\sigma\mu{\bf k}}({\bf r})=\psi_{\eta\sigma\mu}({\bf k}){\rm e}^{i{\bf k}\cdot{\bf r}}$ with $\psi_{\eta\sigma\mu}({\bf k})$ given by
\begin{equation}
\psi_{\eta\sigma\mu}({\bf k})=\frac{1}{\sqrt{2g_{\eta\sigma;k}(g_{\eta\sigma;k}-\mu\lambda_{\eta\sigma})}}\left (
\begin{array}{c}
\eta\mu v_Fk{\rm e}^{-i\eta\varphi_{\bf k}}\\
g_{\eta\sigma;k}-\mu\lambda_{\eta\sigma}
\end{array}
\right ),
\end{equation}
and the corresponding eigenvalue takes the form $\varepsilon_{\eta\sigma\mu}({k})=\mu g_{\eta\sigma;k}$ with $g_{\eta\sigma;k}\equiv \sqrt{v_F^2 k^2+\lambda_{\eta\sigma}^2}$. Here, $k$ and $\varphi_{\bf k}$ are the magnitude and angle of momentum ${\bf k}$, respectively.

It is useful to introduce a unitary transformation $U_{\bf k}\equiv [\psi_{+}({\bf k}),\psi_{-}({\bf k})]$, which corresponds to a change from the pseudospin basis to the pseudohelicity basis. By means of this transformation, Hamiltonian (\ref{H1}) is diagonalized as $\hat h_{\eta\sigma}({\bf k})=U_{\bf k}^+\check h_{\eta\sigma}({\bf k})U_{\bf k}={\rm diag}[\varepsilon_{\eta\sigma;+}(k),\varepsilon_{\eta\sigma;-}(k)]$.

To drive the system out of equilibrium, an electric field ${\bf E}$ is assumed to apply in the $x-y$ plane. In pseudospin basis, this field can be described by a scalar potential, $V=e{\bf E}\cdot {\bf r}$. The portion of observed electric current contributed from electrons with spin index $\sigma$ in $\eta$ valley is determined by ${\bf J}_{\eta\sigma}(T)=-e\sum\limits_{\bf k}{\rm Tr}[\check {\bf j}_{\eta\sigma\bf k}\check \rho_{\eta\sigma}({\bf k},T)]$ with $\check \rho_{\eta\sigma}({\bf k},T)$ as the pseudospin-basis distribution function. $\check {\bf j}_{\eta\sigma\bf k}$ is the single-particle current in pseudospin basis,
$\check {\bf j}_{\eta\sigma\bf k}\equiv -e{\bm \nabla}_{\bf k}\check h_{\eta\sigma}({\bf k})$, which has vanishing diagonal elements: $\check j_{\eta\sigma;x}=ev_F\eta\hat\tau_x$ and $\check j_{\eta\sigma;y}=ev_F\hat\tau_y$. 
By means of the unitary transformation $U_{\bf k}$, ${\bf J}_{\eta\sigma}(T)$ can be determined in pseudo-helicity basis via
\begin{equation}
{\bf J}_{\eta\sigma}(T)=-e\sum_{\bf k}{\rm Tr}[\hat {\bf j}_{\eta\sigma\bf k}\hat \rho_{\eta\sigma}({\bf k},T)]\label{J}
\end{equation}
with $\hat {j}_{\eta\sigma\bf k}=U_{\bf k}^+\check {\bf j}_{\eta\sigma\bf k}U_{\bf k}$ and $\hat \rho_{\eta\sigma}({\bf k},T)=U_{\bf k}^+\check \rho_{\eta\sigma}({\bf k},T)U_{\bf k}$ being the pseudohelicity-basis single-particle current operator and distribution function, respectively. Eq.\,(\ref{J}) can be explicitly written as ${\bf J}_{\eta\sigma}(T)={\bf J}_{\eta\sigma}^{(1)}(T)+{\bf J}_{\eta\sigma}^{(2)}(T)+{\bf J}_{\eta\sigma}^{(3)}(T)$ with
\begin{equation}
{\bf J}_{\eta\sigma}^{(1)}(T)=-v_F^2e\sum_{\bf k}\frac{\bf k}{g_{\eta\sigma;k}}[\hat\rho_{\eta\sigma;++}({\bf k},T)-\hat\rho_{\eta\sigma;--}({\bf k},T)],\label{J1}
\end{equation}
\begin{equation}
{\bf J}_{\eta\sigma}^{(2)}(T)=-2v_Fe\sum_{\bf k}\frac{{\bf k}\lambda_{\eta\sigma}}{kg_{\eta\sigma;k}}{\rm Re}[\hat\rho_{\eta\sigma;+-}({\bf k},T)],\label{J2}
\end{equation} 
and
\begin{equation}
{\bf J}_{\eta\sigma}^{(3)}(T)=-2v_Fe\sum_{\bf k}\eta[{\bf k}\times{\bf n}]{\rm Im}[\hat\rho_{\eta\sigma;+-}({\bf k},T)]/k.\label{J3}
\end{equation} 
Here, ${\bf n}$ is the unit vector along $z$-axis and $\hat\rho_{\eta\sigma;\mu\nu}({\bf k},T)$ [$\mu,\nu=\pm$] are elements of matrix function $\hat\rho_{\eta\sigma}({\bf k},T)$.

To evaluate the current, one has to determine the carrier distribution function. In pseudospin basis it obeys the kinetic equation of the form 
\begin{equation}
\left [\frac{\partial}{\partial T}-e{\bf E}\cdot {\bm \nabla}_{\bf k}\right ]\check \rho_{\eta\sigma}({\bf k},T)+i[\check h_{\eta\sigma}({\bf k}),\check \rho_{\eta\sigma}({\bf k},T)]=-\check I,
\end{equation}
with $\check I$ as the collision term. Applying the unitary transformation, the kinetic equation for pseudohelicity-basis distribution, $\hat \rho_{\eta\sigma}({\bf k},T)$, takes the form (for the sake of brevity, the arguments of distribution function, ${\bf k}$ and $T$, will hereafter be omitted)
\begin{align}
&\left [\frac{\partial}{\partial T}-e{\bf E}\cdot {\bm \nabla}_{\bf k}\right ]\hat \rho_{\eta\sigma}+e{\bf E}\cdot[\hat \rho_{\eta\sigma},U_{\bf k}^+{\bm \nabla}_{\bf k}U_{\bf k}]\nonumber\\
&\,\,\,\,\,\,\,\,\,\,\,\,\,\,\,\,\,\,\,\,+i[\hat h_{\eta\sigma}({\bf k}),\hat \rho_{\eta\sigma}]=-\hat I.\label{KE}
\end{align}
Here, ${\hat I}$ is a scattering term determined by
\begin{eqnarray}
{\hat I}&=& \int \frac{d\omega}{2\pi}[{\hat \Sigma}^r({\bf
k},\omega){\hat {\rm G}}^<({\bf k},\omega)+{\hat \Sigma}^<({\bf
k},\omega){\hat {\rm G}}^a({\bf k},\omega)\nonumber\\
&& - {\hat {\rm G}}^r({\bf k},\omega) {\hat \Sigma}^<({\bf
k},\omega)-{\hat {\rm G}}^<({\bf k},\omega){\hat \Sigma}^a({\bf
k},\omega)],\label{CT}
\end{eqnarray}
with ${\hat {\rm G}}^{r,a,<}({\bf k},\omega)$ and ${\hat
\Sigma}^{r,a,<}({\bf k},\omega)$ as the retarded, advanced and
"lesser" pseudohelicity-basis Green's functions and self-energies,
respectively. Note that in Eqs.\,(\ref{KE}) and (\ref{CT}), the
electron-impurity scattering is embedded in
${\hat \Sigma}^{r,a,<}({\bf k},\omega)$.

Without loss of generality, we further assume that the electric field is applied along the $x$-axis. Thus, the kinetic equation, Eq.\,(\ref{KE}), can be explicitly written as ($\mu=\pm$)
\begin{widetext}
\begin{equation}
\left(\frac{\partial}{\partial T}-e{\bf E}\cdot {\bm \nabla}_{\bf k}\right)\rho_{\eta\sigma;\mu\mu}+\mu eEv_F\left(\frac{\lambda_{\eta\sigma}\cos\varphi_{\bf k}}{g_{\eta\sigma;k}^2}{\rm Re}\rho_{\eta\sigma;+-}+\eta\frac{\sin\varphi_{\bf k}}{g_{\eta\sigma;k}}{\rm Im}\rho_{\eta\sigma;+-}\right)=-\hat I_{\mu\mu}, 
\end{equation}
and
\begin{align}
&\left(\frac{\partial}{\partial T}-e{\bf E}\cdot {\bm \nabla}_{\bf k}\right)\rho_{\eta\sigma;+-}+\frac{eEv_F}{2}\left\{\frac{-1}{g_{\eta\sigma;k}^2}[\lambda_{\eta\sigma}\cos\varphi_{\bf k}+\eta ig_{\eta\sigma;k}\sin\varphi_{\bf k}](\rho_{\eta\sigma;++}-\rho_{\eta\sigma;--})\right .\nonumber\\
&\,\,\,\,\,\,\,\,\,\,\,\,\,\,
\left .-\frac{2i\eta\lambda_{\eta\sigma}\sin\varphi_{\bf k}}{v_Fkg_{\eta\sigma;k}}\rho_{\eta\sigma;+-}\right\}+2ig_{\eta\sigma;k}\rho_{\eta\sigma;+-}=-\hat I_{+-}. 
\end{align}
\end{widetext}

A simplest approach to the complicated collision term $\hat I$ is the relaxation time approximation (RTA). Using it, the kinetic equation in steady-state linear-response regime can be solved analytically. The corresponding results are presented in Appendix. Here, in order to investigate th effects of long-range electron-impurity scattering on longitudinal conductivity as well as on spin- and valley-Hall conductivities, we employ a two-band generalized Kadanoff-Baym ansatz (GKBA) to simplify the collision term $\hat I$.\cite{PhysRevB.34.6933,PSSB:PSSB2221730114} This
ansatz, which expresses the lesser Green's function through the
Wigner distribution function, has been proven sufficiently
accurate to analyze transport and optical properties in
semiconductors.\cite{Haug2008} Further, we consider electron-impurity scattering in the Boltzmann limit, where
the effect of electric field on $\hat {\rm
G}^{r,a}$ is ignored and ${\hat \Sigma}^{r,a}$ and ${\hat \Sigma}^<$ are considered in the self-consistent Born approximation. After complicated but straightforward calculation, $\hat I$ can be explicitly written as
\begin{widetext}
\begin{align}
& \hat I_{\mu\mu}=n_i\sum_{\bf q}|V({\bf k}-{\bf q})|^2\frac{2\pi {\cal D}_{\mu\mu}}{g_{\eta\sigma;k}g_{\eta\sigma;q}}\left \{\left({\cal C}_{\varphi_{\bf k}-\varphi_{\bf q}}kqv_F^2+\lambda_{\eta\sigma}+g_{\eta\sigma;k}g_{\eta\sigma;q}\right )
[\hat \rho_{\eta\sigma;\mu\mu}({\bf k})-\hat \rho_{\eta\sigma;\mu\mu}({\bf q})]\right .\nonumber\\
&\left .+\mu\lambda_{\eta\sigma}v_F(q{\cal C}_{\varphi_{\bf k}-\varphi_{\bf q}}-k){\rm Re}[\hat \rho_{\eta\sigma;+-}({\bf k})]-\mu\lambda_{\eta\sigma}v_F(k{\cal C}_{\varphi_{\bf k}-\varphi_{\bf q}}-q){\rm Re}[\hat \rho_{\eta\sigma;+-}({\bf q})]+\mu\eta{\cal S}_{\varphi_{\bf k}-\varphi_{\bf q}}v_Fkg_{\eta\sigma;q}{\rm Im}[\hat \rho_{\eta\sigma;+-}({\bf q})]\}\right\};
\end{align}
\begin{align}
& \hat I_{+-}=n_i\sum_{{\bf q},\mu}|V({\bf k}-{\bf q})|^2\frac{\pi {\cal D}_{\mu\mu}}{g_{\eta\sigma;k}g_{\eta\sigma;q}}\left \{\mu\left [\lambda_{\eta\sigma}v_F\left(q{\cal C}_{\varphi_{\bf k}-\varphi_{\bf q}}-k\right )+\eta i{\cal S}_{\varphi_{\bf k}-\varphi_{\bf q}}v_Fqg_{\eta\sigma;k}\right ]
[\hat \rho_{\eta\sigma;\mu\mu}({\bf k})-\hat \rho_{\eta\sigma;\mu\mu}({\bf q})]\right .\nonumber\\
&\left .+\left(g_{\eta\sigma;k}g_{\eta\sigma;q}-\lambda_{\eta\sigma}-{\cal C}_{\varphi_{\bf k}-\varphi_{\bf q}}kqv_F^2\right )\left[\hat \rho_{\eta\sigma;+-}({\bf k})+{\cal C}_{\varphi_{\bf k}-\varphi_{\bf q}}\hat \rho_{\eta\sigma;-+}({\bf q})\right ]-kqv_F^2{\cal S}_{\varphi_{\bf k}-\varphi_{\bf q}}^2\hat \rho_{\eta\sigma;-+}({\bf q})\right\}.
\end{align}
\end{widetext}
Here, ${\cal D}_{\mu\mu}\equiv\delta(\varepsilon_{\eta\sigma\mu}(k)-\varepsilon_{\eta\sigma\mu}(q))$, ${\cal S}_{\varphi_{\bf k}-\varphi_{\bf q}}\equiv \sin({\varphi_{\bf k}-\varphi_{\bf q}})$, ${\cal C}_{\varphi_{\bf k}-\varphi_{\bf q}}\equiv \cos({\varphi_{\bf k}-\varphi_{\bf q}})$, $n_i$ is the impurity density, and $V({\bf k}-{\bf q})$ is the electron-impurity scattering potential.

% \section{dc current in the limit of weak electric field}
In present paper, we restrict our consideration to the steady-state linear
response regime. In connection with this, the distribution function can be expressed as a sum of two terms: $ \hat \rho_{\eta\sigma}\approx \hat \rho_{\eta\sigma}^{(0)}+\hat \rho_{\eta\sigma}^{(1)}$,
where $\hat\rho_{\eta\sigma}^{(0)}$ and $\hat\rho_{\eta\sigma}^{(1)}$,
respectively, are the unperturbed part and the linear-electric-field part of $\hat\rho_{\eta\sigma}$. Thus, the kinetic equation for $\hat \rho_{\eta\sigma}^{(1)}({\bf
k})$ can be written as
\begin{equation}
-e{\bf E}\cdot {\bm \nabla}_{\bf k}\hat\rho_{\eta\sigma;\mu\mu}^{(0)}=-\hat I_{\mu\mu}^{(1)}, \label{KE11}
\end{equation}
and
\begin{align}
&\frac{-eEv_F}{2g_{\eta\sigma;k}^2}[\lambda_{\eta\sigma}\cos\varphi_{\bf k}+\eta ig_{\eta\sigma;k}\sin\varphi_{\bf k}](\hat\rho^{(0)}_{\eta\sigma;++}-\hat\rho^{(0)}_{\eta\sigma;--})\nonumber\\
&\,\,\,\,\,\,\,\,\,\,\,\,\,\,
+2ig_{\eta\sigma;k}\hat\rho_{\eta\sigma;+-}^{(1)}=-\hat I_{+-}^{(1)}, \label{KE22}
\end{align}
with $\hat I^{(1)}$ as the linear-electric-field part of the
collision term.

Further, for convenience, we introduce electron and hole distribution functions: $f_{{\rm e};\eta\sigma}({\bf k})\equiv \hat \rho_{\eta\sigma;++}({\bf k})$ and $f_{{\rm h};\eta\sigma}({\bf k})\equiv 1-\rho_{\eta\sigma;--}({\bf k})$. Their unperturbed parts take the forms $f_{{\rm e};\eta\sigma}^{(0)}({\bf k})=n_F(\varepsilon_+(k))$ and $f_{{\rm h};\eta\sigma}^{(0)}=1-n_F(\varepsilon_-(k))$, respectively [$n_F(\varepsilon)=[\exp[(\varepsilon-\mu_0)/(k_0T)]+1]^{-1}$ is the Fermi-Dirac distribution function, $\mu_0$ is the chemical potential, and $T$ is the lattice temperature]. The linear-electric-field parts of $f_{{\rm e};\eta\sigma}$, $f_{{\rm h};\eta\sigma}$ and $\hat \rho_{\eta\sigma;+-}({\bf k})$ can be obtained from Eqs.\,(\ref{KE11}) and (\ref{KE22}) by expanding them in terms of Fourier series of momentum angle: $A^{(1)}({\bf k})=A^{(1,c)}(k)\cos\varphi+A^{(1,s)}(k)\sin\varphi$ with $A$ representing $f_{{\rm e};\eta\sigma}$, $f_{{\rm h};\eta\sigma}$, or $\hat \rho_{\eta\sigma;+-}({\bf k})$. The coefficients of expansion are determined by [$i={\rm e},{\rm h}$]
\begin{widetext}
\begin{equation}
eE\frac{d}{dk}f_{i;\eta\sigma}^{(0)}(k)=\Gamma_1(k)f_{i;\eta\sigma}^{(1,c)}(k)-\Gamma_2(k){\rm Re}[\hat \rho_{\eta\sigma;+-}^{(1,c)}(k)]-\eta\Gamma_3(k){\rm Im}[\hat \rho_{\eta\sigma;+-}^{(1,s)}(k)],\label{EQ1c}
\end{equation} 
\begin{equation}
0=\Gamma_1(k)f_{i;\eta\sigma}^{(1,s)}(k)-\Gamma_2(k){\rm Re}[\hat \rho_{\eta\sigma;+-}^{(1,s)}(k)]+\eta\Gamma_3(k){\rm Im}[\hat \rho_{\eta\sigma;+-}^{(1,c)}(k)],\label{EQ2c}
\end{equation} 
\begin{equation}
eE\frac{v_F\lambda_{\eta\sigma}}{2g_{\eta\sigma;k}^2}\left [\sum_{i={\rm e},{\rm h}}f_{i;\eta\sigma}^{(0)}(k)-1\right ]=-\frac 12 \Gamma_2(k)\sum_{i={\rm e},{\rm h}} f_{i;\eta\sigma}^{(1,c)}(k)+\Gamma_4(k){\rm Re}[\hat \rho_{\eta\sigma;+-}^{(1,c)}(k)]-2g_{\eta\sigma;k}{\rm Im}[\hat \rho_{\eta\sigma;+-}^{(1,c)}(k)],\label{EQ3c}
\end{equation} 
\begin{equation}
0=-\frac 12\Gamma_2(k) \sum_{i={\rm e},{\rm h}} f_{i;\eta\sigma}^{(1,s)}(k)+\Gamma_4(k){\rm Re}[\hat \rho_{\eta\sigma;+-}^{(1,s)}(k)]-2g_{\eta\sigma;k}{\rm Im}[\hat \rho_{\eta\sigma;+-}^{(1,s)}(k)],\label{EQ4c}
\end{equation} 
\begin{equation}
\frac{\eta eEv_F}{2g_{\eta\sigma;k}}\left [\sum_{i={\rm e},{\rm h}}f_{i;\eta\sigma}^{(0)}(k)-1\right ]=-\frac \eta 2\Gamma_3(k) \sum_{i={\rm e},{\rm h}} f_{i;\eta\sigma}^{(1,c)}(k)+\Gamma_4(k){\rm Im}[\hat \rho_{\eta\sigma;+-}^{(1,s)}(k)]+2g_{\eta\sigma;k}{\rm Re}[\hat \rho_{\eta\sigma;+-}^{(1,s)}(k)],\label{EQ5c}
\end{equation} 
and
\begin{equation}
0=\frac \eta 2\Gamma_3(k) \sum_{i={\rm e},{\rm h}} f_{i;\eta\sigma}^{(1,s)}(k)+\Gamma_4(k){\rm Im}[\hat \rho_{\eta\sigma;+-}^{(1,c)}(k)]+2g_{\eta\sigma;k}{\rm Re}[\hat \rho_{\eta\sigma;+-}^{(1,c)}(k)].\label{EQ6c}
\end{equation} 
\end{widetext}
In these equations, the scattering rates $\Gamma_i(k)$ ($i=1,2,3,4$) take the forms 
\begin{equation}
\Gamma_i(k)=n_i\sum_{\bf q}|V({\bf k}-{\bf q})|^22\pi\delta(g_{\eta\sigma;k}-g_{\eta\sigma;q}){\cal B}_{i}({\bf k},{\bf q}),
\end{equation}
where ${\cal B}_1({\bf k},{\bf q})\equiv (1-{\cal C}_{\varphi_{\bf k}-\varphi_{\bf q}})[1+(\lambda_{\eta\sigma}^2+v_F^2kq{\cal C}_{\varphi_{\bf k}-\varphi_{\bf q}})/(g_{\eta\sigma;k}g_{\eta\sigma;q})]$, ${\cal B}_2({\bf k},{\bf q})\equiv (1-{\cal C}_{\varphi_{\bf k}-\varphi_{\bf q}})^2\lambda_{\eta\sigma}v_Fq/(g_{\eta\sigma;k}g_{\eta\sigma;q})$, ${\cal B}_3({\bf k},{\bf q})\equiv {\cal S}_{\varphi_{\bf k}-\varphi_{\bf q}}^2v_Fk/g_{\eta\sigma;q}$, and ${\cal B}_4({\bf k},{\bf q})\equiv [1-\lambda_{\eta\sigma}^2/(g_{\eta\sigma;k}g_{\eta\sigma;q})](1-{\cal C}_{\varphi_{\bf k}-\varphi_{\bf q}}^2)$.

After the coefficients of Fourier series are determined, the charge current contributed from electrons with spin $\sigma$ in $\eta$ valley, ${\bf J}_{\eta\sigma}$, can be obtained via
\begin{widetext}
\begin{equation}
{J}_{\eta\sigma;x}=-v_Fe\sum_{\bf k}\left\{\frac{v_Fk}{2g_{\eta\sigma;k}}\sum_{i={\rm e},{\rm h}}f^{(1,c)}_{i;\eta\sigma}(k)+\frac{\lambda_{\eta\sigma}}{g_{\eta\sigma;k}}{\rm Re}[\hat\rho^{(1,c)}_{\eta\sigma;+-}(k)]
+\eta{\rm Im}[\hat\rho^{(1,s)}_{\eta\sigma;+-}(k)]\right\},\label{Jx}
\end{equation} 
and
\begin{equation}
{J}_{\eta\sigma;y}=-v_Fe\sum_{\bf k}\left\{\frac{v_Fk}{2g_{\eta\sigma;k}}\sum_{i={\rm e},{\rm h}}f^{(1,s)}_{i;\eta\sigma}(k)+\frac{\lambda_{\eta\sigma}}{g_{\eta\sigma;k}}{\rm Re}[\hat\rho^{(1,s)}_{\eta\sigma;+-}(k)]
-\eta{\rm Im}[\hat\rho^{(1,c)}_{\eta\sigma;+-}(k)]\right\}.\label{Jy}
\end{equation} 
\end{widetext}
From Eq.\,(\ref{Jx}) it is clear that not only the nonequilibrium distribution of electrons or holes makes nonvanishing contribution to ${J}_{\eta\sigma;x}$, the interband correlation process induced by dc electric field also contributes to the longitudinal current. Due to this interband coherence, zero-temperature ${J}_{\eta\sigma;x}$ remains finite when chemical potential lies within the energy gap. Another interesting property of ${J}_{\eta\sigma;x}$ is the symmetric relation: ${J}_{++;x}={J}_{--;x}$ and ${J}_{+-;x}={J}_{-+;x}$, {\it i.e.} the total current of electrons with both spins in $K$ node and that in $K'$ node are the same.   

From Eq.\,(\ref{Jy}) we see that the transverse current $J_{\eta\sigma;y}$ is nonvanishing although the total Hall current $J_y=\sum\limits_{\eta,\sigma}J_{\eta\sigma;y}$ is zero. The nonvanishing of $J_{\eta\sigma;y}$ not only directly comes from the interband coherence, {\it i.e.} from nonvanishing of ${\rm Re}[\hat\rho^{(1,s)}_{\eta\sigma;+-}(k)]$ and ${\rm Im}[\hat\rho^{(1,c)}_{\eta\sigma;+-}(k)]$, it is also associated with the nonvanishing of $f^{(1,s)}_{i;\eta\sigma}(k)$ arising from an impurity-mediated process: due to electron-impurity scattering, the coherence process between conduction and valence bands causes electrons or holes moving transversely [this can be seen from Eq.\,(\ref{EQ2c})]. Note that in the formalism of transport within relaxation-time approximation presented in Appendix,  $f^{(1,s)}_{i;\eta\sigma}(k)$ is completely absent. 

It is interesting to analyze the impurity-density dependencies of $J_{\eta\sigma;x}$ and $J_{\eta\sigma;y}$ in the limit $n_i\to 0$. From Eqs.\,(\ref{EQ1c})-(\ref{EQ6c}) it follows that when $n_i\to 0$, $f_{i;\eta\sigma}^{(1,s)}$, ${\rm Re}\rho_{\eta\sigma}^{(1,s)}$, and ${\rm Im}\rho_{\eta\sigma}^{(1,c)}$ are essentially independent of $n_i$, while the dependencies of $f_{i;\eta\sigma}^{(1,c)}$, ${\rm Re}\rho_{\eta\sigma}^{(1,c)}$, and ${\rm Im}\rho_{\eta\sigma}^{(1,s)}$ on $n_i$ are quite complicated: the Laurent series of each of these functions in the case $n_i\to 0$ contains two terms, one is proportional to $n_i$ and another one is inversely proportional to $n_i$. Consequently, when $n_i\to 0$, $J_{\eta\sigma;y}$ is independent of $n_i$ but $J_{\eta\sigma;x}$ contains two components, $J_{\eta\sigma;x}\to Cn_i+D/n_i$. 

\section{Numerical results}

Further, considering finite-temperature screened electron-impurity scattering, we present a numerical calculation to investigate the longitudinal transport and spin- and valley-Hall effects. Whence linear system of Eqs.\,(\ref{EQ1c})-(\ref{EQ6c}) is solved, the currents $J_{\eta\sigma;x}$ and $J_{\eta\sigma;y}$ are determined. Thus, we can obtain the total longitudinal, spin- and valley-Hall conductivities, $\sigma_{xx}$, $\sigma_{yx}^{(s)}$, and $\sigma_{yx}^{(v)}$, defined as $\sigma_{xx}=J_x/E$,
$\sigma_{yx}^{(s)}=\sum\limits_{\sigma,\eta;i=1,2,3} \sigma J_{\eta\sigma;y}^{(i)}/(-2eE)$ and $\sigma_{yx}^{(v)}=\sum\limits_{\sigma,\eta;i=1,2,3} \eta J_{\eta\sigma;y}^{(i)}/(-2eE)$, respectively. In calculation, a screened scattering potential due to charged impurities is considered: $V(q)=e^2/[2\kappa\epsilon_0(q+q_s)\epsilon(q)]$ with $\kappa$ as the dielectric constant of substrate. Here, to overcome the divergence of scattering potential in the limit of $q\to 0$ when $\mu_0$ lies within energy gap, a formal cut-off parameter $\Lambda\equiv v_Fq_s$ is introduced.  $\epsilon(q)=1-v(q)\Pi(q,T,\mu_0)$ is the static dielectric function, $v(q)=e^2/(2\epsilon_0q)$ is the 2D Coulomb potential, and $\Pi(q,T,\mu_0)$ is the static polarization function for chemical potential $\mu_0$ at finite temperature $T$. At zero temperature $\Pi(q,T=0,\mu_0)$ takes the form\cite{PhysRevD.33.3704,PhysRevB.78.075433,0953-8984-21-7-075303,0953-8984-21-2-025506,1402-4896-83-3-035002,PhysRevB.89.195410,PhysRevB.90.035142}
\begin{align*}
\Pi(q,T=0,\mu_0)&=-\frac{\mu_0}{2\pi v_F^2}\sum_{\sigma,\eta}\left [F(q)\theta(|\lambda_{\eta\sigma}|-\mu_0)\right .\nonumber\\
&\left . +G(q)\theta(\mu_0-|\lambda_{\eta\sigma}|)\right],
\end{align*} 
with $F(q)$ and $G(q)$, respectively, taking the forms [$k_F^{\eta\sigma}=\sqrt{\mu_0^2-\lambda_{\eta\sigma}^2}$]
\begin{equation}
F(q)=\frac{|\lambda_{\eta\sigma}|}{2\mu_0}+\frac{v_F^2q^2-4\lambda_{\eta\sigma}^2}{4v_Fq\mu_0}\arcsin\left(\sqrt{\frac{v_F^2q^2}{v_F^2q^2+4\lambda_{\eta\sigma}^2}}\right)
\end{equation}
and
\begin{align*}
&G(q)=1-\theta(q-2k_F^{\eta\sigma})\left[\frac{\sqrt{q^2-4(k_F^{\eta\sigma})^2}}{2q}\right .\nonumber\\
&\left .-\frac{v_F^2q^2-4\lambda_{\eta\sigma}^2}{4v_Fq\mu_0}\arctan\left({\frac{\sqrt{v_F^2q^2-4(k_F^{\eta\sigma})^2v_F^2}}{2\mu_0}}\right)\right ].
\end{align*}
The finite-temperature static polarization function is determined via
\begin{equation}
\Pi(q,T,\mu_0)=\frac{1}{4T}\int_{-\infty}^{\infty}d\varepsilon\frac{\Pi(q,T=0,\varepsilon)}{\cosh^2[(\mu_0-\epsilon)/(2T)]}.
\end{equation}

\subsection{Longitudinal conductivity}

\begin{figure}
\includegraphics[width=0.45\textwidth,clip] {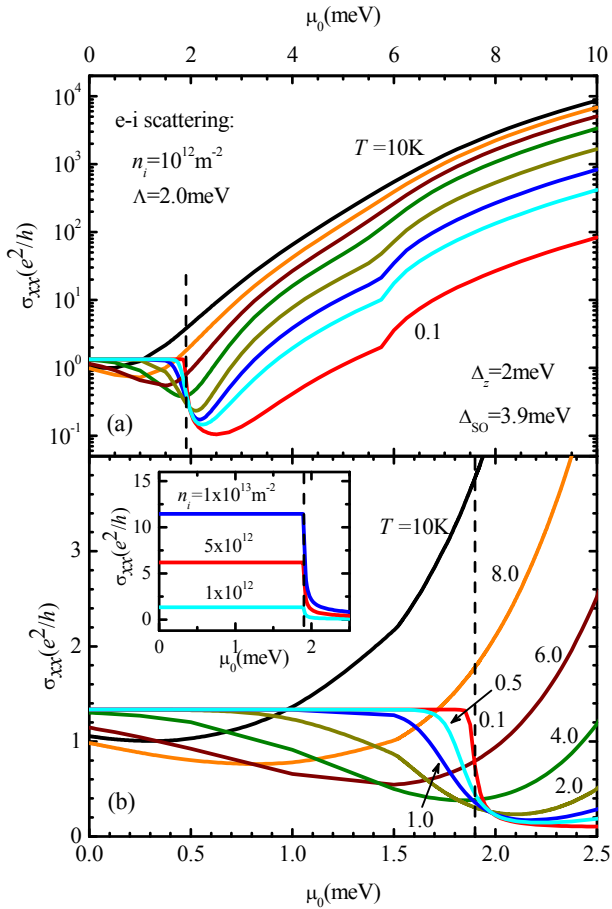}
\caption{(Color online) (a) Dependencies of longitudinal conductivity on chemical potential 
at various temperatures. From top to bottom, the lattice temperatures are  $T=10.0$,
  $8.0$, $6.0$, $4.0$, $1.0$, $0.5$, and
  $0.1$\,K. (b) Enlarged version of (a) for chemical potential lying within energy gap and near the bottom of conduction band.
In these figures, $\Delta_{\rm SO}$=3.9\,meV, $\Delta_z=2.0$\,meV, and hence the bottom of conduction band is at the energy $1.9$\,meV, which is denoted by the vertical dash lines. The impurity density is assumed to be $n_{i}=1\times
10^{-12}$\,m$^{-2}$ and cut-off parameter is chosen to be $\Lambda=2$\,meV. In inset of (b) $\sigma_{xx}$ versus $\mu_0$ at lattice temperature $T=0.001$\,K are plotted for various impurity densities: $n_i=1\times 10^{12}$, $5\times 10^{12}$, and $1\times 10^{13}$\,m$^{-2}$.} \label{fig1}
\end{figure}

In Fig.\,1, we plot the dependencies of longitudinal conductivity on chemical potential at various temperature. We see that when chemical potential decreases but still lies above the bottom of conduction band, the conductivity decreases rapidly. This is due to the fact that the density of carriers excited by temperature exponentially decreases. However, when $\mu_0$ decreases to near the bottom of conduction band and further drops into the energy gap [see Fig.\,1(b)], the behaviors of $\sigma_{xx}$ versus $\mu_0$ at relatively low temperature and at high temperature are quite different.  At low temperature, the conductivity first decreases and then increases with a decrease of chemical potential. When chemical potential is close to the center of the energy gap, the conductivity finally reaches a constant value of order of $e^2/h$, forming a plateau. However, when temperature increases, the width of this plateau becomes shorter. It even disappears at relatively high temperature $k_0T>||\Delta_z|-\Delta_{\rm SO}|$:  in this case, the conductivity first decreases and then increases when $\mu_0$ decreases. Note that if the chemical potential $\mu_0$ ascends from the bottom of conduction band and goes close to the value $|\Delta_z|+\Delta_{\rm SO}$, the second conduction band is no longer empty: gradual change of $\sigma_{xx}$ with $\mu_0$ is broken and "dog-leg" shaped connection can be observed near $\mu_0\approx \Delta_z+\Delta_{\rm SO}=3.9$\,meV at relatively low temperatures, {\it i.e.} at $T=0.1$, $0.5$, $1.0$\,K (see Fig.\,1(a)). 

\begin{figure}
\includegraphics [width=0.45\textwidth,clip] {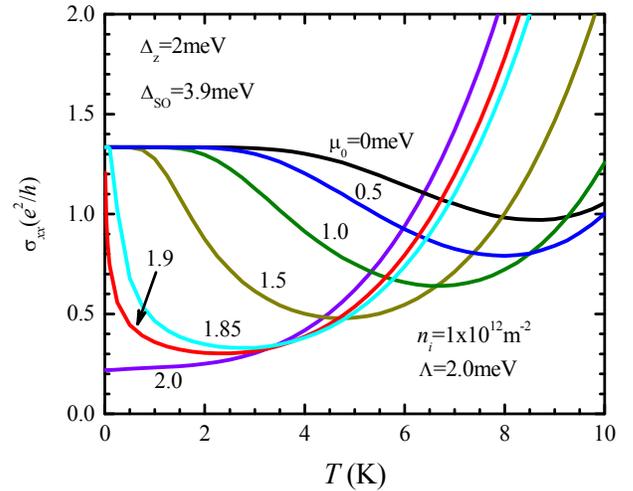}
\caption{(Color online) Dependencies of longitudinal conductivity on temperature for various
chemical potentials: $\mu_0$=0, 0.5, 1.0, 1.5, 1.85, 1.9, and 2.0\,meV. Other parameters are the same as in Fig.\,1. } \label{fig2}
\end{figure}

Obviously, the low-temperature plateau values in $\sigma_{xx}$ versus $\mu_0$ depend on the strength of electron-impurity scattering. To show this, in the inset of Fig.\,1(b) we plot the dependencies of conductivity on chemical potential for various impurity densities at temperature $T=0.001$\,K. It can be seen that the values of residual conductivity increase when the strength of electron-impurity ascends. Note that our approach is valid only for relatively clean samples ({\it i.e.} for small $n_i$). If impurity density further increases, one has to consider the collisional broadening effect induced by electron-impurity interaction on transport, which is beyond the scope of present study.

\begin{figure}
\includegraphics [width=0.45\textwidth,clip] {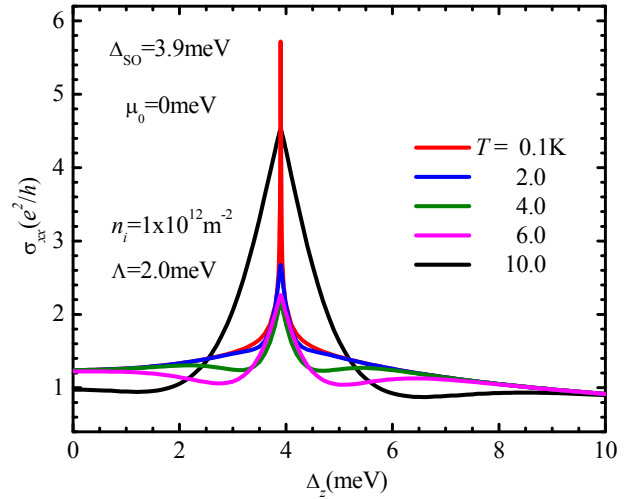}
\caption{(Color online) $\sigma_{xx}$ versus $\Delta_z$ at various
lattice temperatures: $T$=0.1, 2, 4, 6, and 10\,K. Other parameters are the same as in Fig.\,1. } \label{fig3}
\end{figure}

When a chemical potential increases across the bottom of conduction band, the metal-insulator transition (MIT) is expected to be observed. To demonstrate this, in Fig.\,2 we plot the dependencies of longitudinal conductivity on temperature for various chemical potentials. We find that when temperature increases, the conductivity first decreases from the residual-conductivity value and then increases when $\mu_0$ lies within the energy gap, while it monotonically increases in the case $\mu_0>||\Delta_z|-\Delta_{\rm SO}|$. Note that the observed descent of $\sigma_{xx}$ with an ascent of $T$ for $\mu_0<||\Delta_z|-\Delta_{\rm SO}|$ is associated with temperature dependence of screening of electron-impurity scattering. In previous studies on MIT in conventional 2D electron systems, finite-temperature screening plays a key role in mechanism proposed by Das Sarma and Hwang.\cite{PhysRevLett.83.164,PhysRevB.68.195315,PhysRevB.69.195305} In Fig.\,2, we also can see that when $\mu_0$ lies within the energy gap and it is far from the bottom of conduction band (in our case, $\mu_0$=0, 0.5, 1.0, 1.5\,meV), a plateau is formed in $\sigma_{xx}$ versus $T$ when temperature increases from zero. This implies that for such chemical potentials and lattice temperatures, the interband correlation makes dominant contribution to total conductivity. When chemical potential further ascends and is close to the bottom of conduction bands, the plateau becomes smaller and finally disappears.

In silicene, due to the specific buckled structure, $\Delta_z$ can be tunned by applying an electric field perpendicular to the plane of atoms. In Fig.\,3, we plot the dependencies of longitudinal conductivity on $\Delta_z$  at various lattice temperatures in the case $\mu_0=0$. We can see that when $\Delta_z$ goes close to $\Delta_{\rm SO}$ from both the left and right sides, ({\it i.e.} the energy gap decreases), conductivity monotonically increases at relatively low temperature ($T=0.1$ and 2\,K) while it first decreases and then increases at high temperature (at $T=4$, 6, and 10\,K).  
 
\begin{figure}
\includegraphics [width=0.45\textwidth,clip] {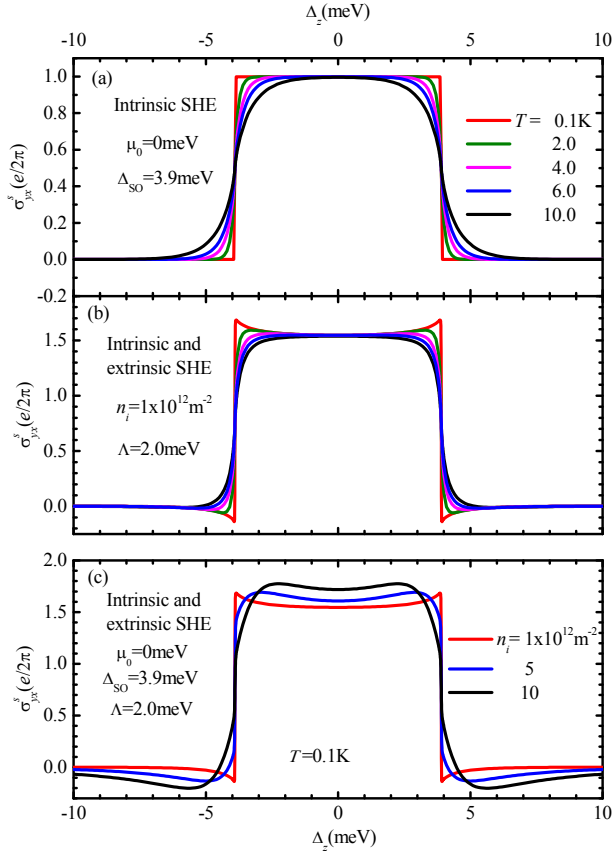}
\caption{(Color online) Intrinsic (a) as well as intrinsic and extrinsic (b) spin-Hall conductivities versus $\Delta_z$ at various
lattice temperatures: $T$=0.1, 2, 4, 6, and 10\,K. (c) $\sigma_{xy}^{(s)}$ versus $\Delta_z$ at $T=0.1$\,K for various impurity densities: $n_i=1\times 10^{12}$, $5\times 10^{12}$, and $1\times 10^{13}$\,m$^{-2}$. Other parameters in (a)-(c) are the same as in Fig.\,1. } \label{fig4}
\end{figure}

\begin{figure}
\includegraphics [width=0.45\textwidth,clip] {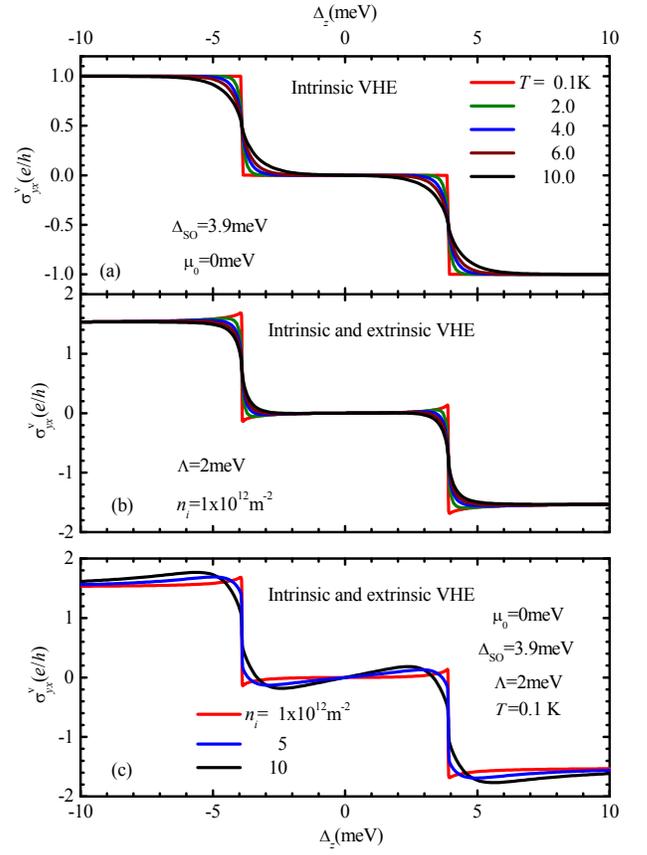}
\caption{(Color online) Intrinsic (a) as well as intrinsic and extrinsic (b) valley-Hall conductivities versus $\Delta_z$ at various
lattice temperatures. (c) $\sigma_{xy}^{(v)}$ versus $\Delta_z$ at $T=0.1$\,K for various impurity densities. Other parameters in (a)-(c) are the same as in Fig.\,1. } \label{fig5}
\end{figure}

\begin{figure}
\includegraphics [width=0.45\textwidth,clip] {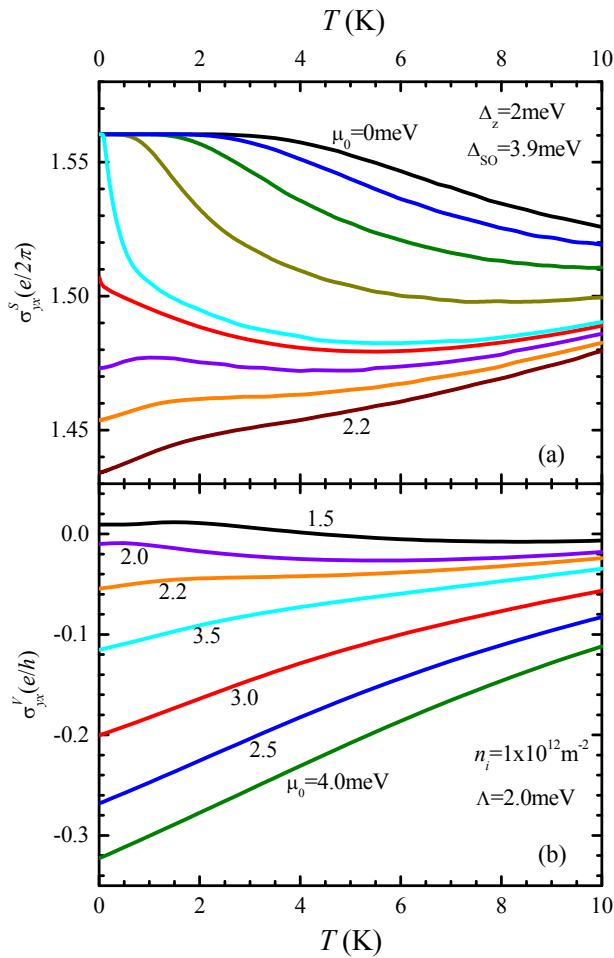}
\caption{(Color online) Temperature dependencies of total spin- (a) and valley-Hall (b) conductivities for various chemical potentials. In (a), from top to bottom, the chemical potentials are $\mu_0=0$, $0.5$, $1.0$, $1.5$, $1.85$, $1.9$, $2.0$, $2.1$, and $2.2$\,meV, respectively. Other parameters in (a) and (b) are the same as in Fig.\,1. } \label{fig6}
\end{figure}

\subsection{Spin- and Valley-Hall effects}

When $|\Delta_z|$ changes across $\Delta_{\rm SO}$, the topological phase transition may occur and the spin- and valley-Hall conductivities are expected to change abruptly. In Figs.\,4 and 5, we plot the dependencies of SHC and VHC on $\Delta_{z}$ at various temperatures. From Figs.\,4(a) and 5(a), it is clear that at low temperature, the intrinsic SHC is nonvanishing only when $|\Delta_{z}|<\Delta_{\rm SO}$, while the intrinsic VHC is nonvanishing only in the case $|\Delta_{z}|>\Delta_{\rm SO}$ and its sign changes with the change of sign of $\Delta_z$. Their nonvanishing values are universal: $\sigma_{yx}^{(s)}=e/(2\pi)$ for $|\Delta_{z}|<\Delta_{\rm SO}$; $|\sigma_{yx}^{(v)}|=e/h$ for $|\Delta_{z}|>\Delta_{\rm SO}$. When temperature increases, the sharp changes of $\sigma_{yx}^{(s)}$ and $\sigma_{yx}^{(v)}$ near $|\Delta_z|=\Delta_{\rm SO}$ are smeared out. From Figs.\,4(b) and 5(b), we see that the electron-impurity scattering makes additional contribution to the total $\sigma_{yx}^{(s)}$ and $\sigma_{yx}^{(v)}$: the plateau values of $\sigma_{yx}^{(s)}$ and of $\sigma_{yx}^{(v)}$, respectively, are $\sigma_{yx}^{(s)}\approx 1.67e/(2\pi)$ (for $|\Delta_{z}|<\Delta_{\rm SO}$) and $|\sigma_{yx}^{(v)}|\approx 1.53e/h$ (for $|\Delta_{z}|>\Delta_{\rm SO}$). At the same time, differing from the intrinsic SHC and VHC, $\sigma_{yx}^{(s)}$ no longer vanishes when $|\Delta_{z}|>\Delta_{\rm SO}$ and $\sigma_{yx}^{(v)}$ is also finite in the case $|\Delta_{z}|<\Delta_{\rm SO}$ at relatively low temperature:  when $|\Delta_z|$ increases from the case $|\Delta_{z}|<\Delta_{\rm SO}$ to the case $|\Delta_{z}|>\Delta_{\rm SO}$, $\sigma_{yx}^{(s)}$ first increases from the plateau value, quickly falls to a negative value and then increases towards zero. Such peak-dip structure also can be observed in $\sigma_{yx}^{(v)}$ versus $\Delta_z$ near $|\Delta_{z}|=\Delta_{\rm SO}$. When temperature ascends, the peak-dip features in $\Delta_z$-dependencies of $\sigma_{yx}^{(s)}$ and $\sigma_{yx}^{(v)}$ are gradually smeared out and finally disappear, but the drastic change of values near phase-transition point still can be observed.      

Obviously, the plateau values of spin- and valley-Hall conductivity are no longer universal in the presence of electron-impurity scattering. In Figs.\,4(c) and 5(c), we plot the total spin- and valley-Hall conductivities versus $\Delta_z$ for various impurity density. It is clear that although the electron-impurity scattering makes substantial contribution to spin- and valley-Hall conductivities, the values of total $\sigma_{yx}^{(s)}$ and $\sigma_{yx}^{(v)}$ slightly depend on the impurity density. This can be seen from the fact that in the limit of $n_i\to 0$, $f_{\eta\sigma}^{(1,s)}$, ${\rm Re}\rho_{\eta\sigma}^{(1,s)}$, and ${\rm Im}\rho_{\eta\sigma}^{(1,c)}$ is essentially independent of impurity density.

It is interesting to analyze the temperature dependencies of total spin- and valley-Hall conductivities for various chemical potentials, which are plotted in Fig.\,6. When temperature increases,  the spin-Hall conductivity decreases in the case $|\mu_0|>||\Delta_z|-\Delta_{\rm SO}|$ while it increases for $|\mu_0|<||\Delta_z|-\Delta_{\rm SO}|$, exhibiting a crossover behavior due to "metal-insulator transition". In contrast to this, the valley-Hall conductivity monotonically increases with an increase of temperature for $\mu_0>||\Delta_z|-\Delta_{\rm SO}|$. In the case $|\mu_0|$ increases across the bottom of conduction band, it rapidly reduces towards zero.  

\section{Conclusions}

Linear longitudinal transport as well as spin- and valley-Hall effects in silicene have been investigated by means of a pseudohelicity-basis kinetic equation approach. A numerical calculation was carried out by considering finite-temperature-screened electron-impurity scattering. We found that, when the density of equilibrium carriers vanishes, the low-temperature longitudinal conductivity still is nonvanishing. This residual conductivity strongly depends on the electron-impurity scattering and leads to a plateau in the dependence of conductivity on chemical potential at relatively low temperature. We also clarified that the electron-impurity interaction makes substantial contribution to the total spin- and valley-Hall conductivities, although the values of these condictivities are almost independent of impurity density. The temperature dependencies of longitudinal conductivity and of spin- and valley-Hall conductivities for various chemical potentials were also carried out. We found that the changes of $\sigma_{xx}$ and $\sigma_{yx}^{(s)}$ with temperature are quite different for chemical potentials lying below and above the bottom of conduction band, and they exhibit crossover behaviors due to metal-insulator transition.

\begin{acknowledgments}
This work was supported by the project of National Key Basic Research
Program of China (973 Program) (Grant No. 2012CB927403) and National
Natural Science Foundation of China (Grant Nos. 11274227 and 91121021). 
\end{acknowledgments}

\appendix*

\section{Steady-state linear transport within relaxation time approximation}

In the relaxation time approximation, the scattering term can be written as $\hat I_{\mu\nu}=\Gamma_{\mu\nu} (\rho_{\eta\sigma;\mu\nu}-\rho_{\eta\sigma;\mu\nu}^{(0)})$ with $\Gamma_{\mu\nu}$ as the formal parameters of scattering rates. The elements of unperturbed distribution function, $\rho_{\eta\sigma;\mu\nu}^{(0)}$, take the forms:
$\hat\rho_{\eta\sigma;\mu\mu}^{(0)}=n_F(\varepsilon_{\eta\sigma\mu}(k))$ and $\hat\rho_{\eta\sigma;\mu\nu}^{(0)}=0$ if $\mu\neq\nu$.

From Eqs.\,(\ref{KE11}) and (\ref{KE22}) it follows that $\hat\rho_{\eta\sigma;+-}^{(1)}$ takes the form
\begin{align*}
&\hat\rho_{\eta\sigma;+-}^{(1)}({\bf k}) = 
\frac{eEv_F(\lambda_{\eta\sigma}\cos\varphi_{\bf k}+\eta ig_{\eta\sigma;k}\sin\varphi_{\bf k})}{2g_{\eta\sigma;k}^2(\Gamma_{+-}+2ig_{\eta\sigma;k})}\nonumber\\
&\,\,\,\,\,\,\,\,\,\,\,\,\,\,\times[\hat\rho_{\eta\sigma;++}^{(0)}(k)-\hat\rho_{\eta\sigma;--}^{(0)}(k)],
\end{align*} 
and $\hat \rho^{(1)}_{\eta\sigma;\mu\mu}({\bf k})$ are given by
\begin{equation}
\hat \rho_{\eta\sigma;\mu\mu}^{(1)}({\bf k})= \frac{eE\cos\varphi_{\bf k}}{\Gamma_{\mu\mu}}\frac{d}{d k} \hat \rho^{(0)}_{\eta\sigma;\mu\mu}(k).
\end{equation}
Hence, the components of longitudinal current $J_{\eta\sigma;x}$, $J_{\eta\sigma;x}^{(i)}$ ($i=1,2,3$),  are determined via 
\begin{equation}
J_{\eta\sigma;x}^{(1)}=- v_F^4e^2E\sum_{{\bf k},\mu}\frac{k^2}{2\Gamma_{\mu\mu}g_{\eta\sigma;k}^2}\frac{\partial \hat \rho^{(0)}_{\eta\sigma;\mu\mu}(k)}{\partial \varepsilon_{\eta\sigma\mu}(k)},\label{J1x}
\end{equation} 
\begin{equation}
J_{\eta\sigma;x}^{(2)}=-e^2E\lambda_{\eta\sigma}^2 v_F^2\Gamma_{+-}\sum_{\bf k}\frac{\hat\rho_{\eta\sigma;++}^{(0)}(k)-\hat\rho_{\eta\sigma;--}^{(0)}(k)}{2g_{\eta\sigma;k}^3(\Gamma_{+-}^2+4g_{\eta\sigma;k}^2)},\label{J2x}
\end{equation}
and
\begin{equation}
J_{\eta\sigma;x}^{(3)}=-e^2Ev_F^2\Gamma_{+-}\sum_{\bf k}\frac{[\hat\rho_{\eta\sigma;++}^{(0)}(k)-\hat\rho_{\eta\sigma;--}^{(0)}(k)]}{2g_{\eta\sigma;k}[\Gamma_{+-}^2+4g_{\eta\sigma;k}^2]}.\label{J3x}
\end{equation}
Obviously, $J_{\eta\sigma;y}^{(1)}=0$ but $J_{\eta\sigma;y}^{(2)}$ and $J_{\eta\sigma;y}^{(3)}$ are finite due to nonvanishing of $\hat \rho_{\eta\sigma;+-}$:
\begin{equation}
J_{\eta\sigma;y}^{(2)}=J_{\eta\sigma;y}^{(3)}=-\eta e^2E\lambda_{\eta\sigma} v_F^2\sum_{\bf k}\frac{\hat\rho_{\eta\sigma;++}^{(0)}(k)-\hat\rho_{\eta\sigma;--}^{(0)}(k)}{g_{\eta\sigma;k}(\Gamma_{+-}^2+4g_{\eta\sigma;k}^2)}.\label{J2y}
\end{equation}

From Eqs.\,(\ref{J2x}) and (\ref{J3x}) it is clear that the interband correlation makes nonvanishing contribution to longitudinal current, which is proportional to $\Gamma_{+-}$, {\it i.e.} the scattering rate. This is significantly different from the contribution to current from the nonequilibrium electrons and holes [Eq.\,(\ref{J1x})], which is proportional to the relaxation time, {\it i.e.}, $1/\Gamma_{\mu\mu}$. 

Another interesting result obtained from Eqs.\,(\ref{J2x}) and (\ref{J3x}) is that $J_{2x}$ and $J_{3x}$ remain nonvanishing even when the equilibrium densities of electrons and holes are zero. To carry out this residual conductivity, we consider the case in which the lattice temperature tends to zero and the chemical potential lies within the energy gap: $|\mu_0|<||\Delta_z|-\Delta_{\rm SO}|$. Obviously, $J_{\eta\sigma;x}^{(1)}$ vanishes since there is no electron near the Fermi surface: $\frac{\partial \hat \rho^{(0)}_{\eta\sigma;\mu\mu}(k)}{\partial \varepsilon_{\mu k}}=-\delta(\varepsilon_{\eta\sigma\mu}(k)-\mu_0)\rightarrow 0 $. However, $J_{\eta\sigma;x}^{(2)}$ and $J_{\eta\sigma;x}^{(3)}$ are nonvanishing and take finite values. Using the fact that $\hat\rho_{\eta\sigma;++}^{(0)}(k)-\hat\rho_{\eta\sigma;--}^{(0)}(k)=-1$ at $T=0$, we obtain the conductivity associated with $J_{\eta\sigma;x}^{(2)}$ and $J_{\eta\sigma;x}^{(3)}$ ($\sigma_{xx}^{(2)}\equiv \sum\limits_{\eta,\sigma}J_{\eta\sigma;x}^{(2)}/E$, $\sigma_{xx}^{(3)}\equiv \sum\limits_{\eta,\sigma}J_{\eta\sigma;x}^{(3)}/E$)
\begin{eqnarray}
\left. \sigma_{xx}^{(2)}\right|_{T=0}&=&\sum_{\eta,\sigma}\frac{v_F^2\lambda_{\eta\sigma}^2\Gamma_{+-} e^2}{4\pi}\int_0^\infty \frac{kdk}{g_{\eta\sigma;k}^3[\Gamma_{+-}^2+4g_{\eta\sigma;k}^2]}\nonumber\\
&=&\sum_{\eta\sigma}\frac{e^2|\lambda_{\eta\sigma}|}{4\pi\Gamma_{+-}^2}\left\{|\lambda_{\eta\sigma}|[2\arctan(2|\lambda_{\eta\sigma}|/\Gamma_{+-})-\pi]\right.\nonumber\\
&&\left .+\Gamma_{+-}\right\},\label{S1}
\end{eqnarray}   
and
\begin{eqnarray}
\left .\sigma_{xx}^{(3)}\right |_{T=0}&=&\sum_{\eta,\sigma}\frac{v_F^2\Gamma_{+-} e^2}{4\pi}\int_0^\infty \frac{kdk}{g_{\eta\sigma;k}[\Gamma_{+-}^2+4g_{\eta\sigma;k}^2]}\nonumber\\
&=&\frac{e^2}{16\pi}\sum_{\eta,\sigma}[\pi-2\arctan(2|\lambda_{\eta\sigma}|/\Gamma_{+-})].\label{S2}
\end{eqnarray}

From Eqs.\,(\ref{S1}) and (\ref{S2}) we clarify that the residual conductivity depends on the factor $\Gamma_{+-}/\lambda_{\eta\sigma}$. In the case of small $\Gamma_{+-}/\lambda_{\eta\sigma}$ for any $\eta$ and $\sigma$, we have $\left.\sigma_{xx}^{(2)}\right|_{T=0}\approx \sum\limits_{\eta,\sigma}e^2\Gamma_{+-}/(48\pi|\lambda_{\eta\sigma}|)$
and $\left .\sigma_{xx}^{(3)}\right|_{T=0}\approx \sum\limits_{\eta,\sigma} e^2\Gamma_{+-}/(16\pi|\lambda_{\eta\sigma}|)$. Thus, the total residual conductivity takes the form $\left .\sigma_{xx}\right|_{T=0}=\left .\sigma_{xx}^{(2)}\right|_{T=0}+\left .\sigma_{xx}^{(3)}\right|_{T=0}=\sum\limits_{\eta,\sigma} e^2\Gamma_{+-}/(12\pi|\lambda_{\eta\sigma}|)$, which depends on scattering and on $\lambda_{\eta\sigma}$. In the opposite limit, {\it i.e.} in the case of small $\lambda_{\eta\sigma}/\Gamma_{+-}$ for any $\eta$ and $\sigma$, $\left .\sigma_{xx}^{(2)}\right|_{T=0}\approx 0$ and $\left .\sigma_{xx}^{(3)}\right|_{T=0}\approx e^2/4=(\pi/8)(e^2/h) $ reduces to a universal value. Note that in the previous investigation on transport of zero-gap Dirac fermions, various values of conductivity have been obtained.\cite{RevModPhys.83.407}

It is interesting to analyze the spin- and valley-Hall conductivities within relaxation time approximation.  From (\ref{J2y}) we find that, when $\mu_0$ lies within the energy gap, zero-temperature $\sigma_{yx}^{(s)}$ and $\sigma_{yx}^{(v)}$ take the forms
\begin{equation}
\left .\sigma_{yx}^{(s)}\right |_{T=0}=-e\sum_{\eta,\sigma}\eta\sigma \frac{\lambda_{\eta\sigma}}{8\pi\Gamma_{+-}}[\pi-2\arctan(2|\lambda_{\eta\sigma}|/\Gamma_{+-})]
\end{equation}
and
\begin{equation}
\left .\sigma_{yx}^{(v)}\right|_{T=0}=-e\sum_{\eta,\sigma} \frac{\lambda_{\eta\sigma}}{8\pi\Gamma_{+-}}[\pi-2\arctan(2|\lambda_{\eta\sigma}|/\Gamma_{+-})]
\end{equation}
respectively. In the clean limit $\Gamma_{+-}/|\lambda_{\eta\sigma}|\rightarrow 0$, spin-Hall conductivity reduces to $\left .\sigma_{yx}^{(s)}\right|_{T=0}=-e/(8\pi)\sum_{\eta,\sigma}\eta\sigma \lambda_{\eta\sigma}/|\lambda_{\eta\sigma}|$, which takes the value
\begin{equation}
\left.\sigma_{yx}^{(s)}\right|_{T= 0}\rightarrow
\left\{
\begin{array}{cc}
e/(2\pi)&\,\,\,\,\,|\Delta_z|<\Delta_{\rm SO}\\
0&\,\,\,\,\,|\Delta_z|>\Delta_{\rm SO}
\end{array}
\right ..
\end{equation}
In the same way, $\left .\sigma_{yx}^{(v)}\right|_{T=0}$ can be obtained when $\Gamma_{+-}/|\lambda_{\eta\sigma}|\rightarrow 0$: 
\begin{eqnarray}
\left .\sigma_{yx}^{(v)}\right|_{T=0}&=&-e/(8\pi)\sum_{\eta,\sigma}\lambda_{\eta\sigma}/|\lambda_{\eta\sigma}|\nonumber\\
&=&\left\{
\begin{array}{cc}
-e/(2\pi)&\,\,\,\,\,\Delta_z>\Delta_{\rm SO}\\
0&\,\,\,\,\,|\Delta_z|<\Delta_{\rm SO}\\
e/(2\pi)&\,\,\,\,\,\Delta_z<-\Delta_{\rm SO}
\end{array}
\right ..
\end{eqnarray}
These results agree with the previous studies: when the values of $|\Delta_z|$ changes
from $|\Delta_z|<\Delta_{\rm SO}$ to $|\Delta_z|>\Delta_{\rm SO}$, the spin-Hall effect vanishes while valley-Hall effect appears.\cite{APL.102.162412,PhysRevB.87.235426,0953-8984-26-34-345303}

% \section{Conclusions}
\bibliography{transport}
%\begin{thebibliography}{100}
%\end{thebibliography}
\end{document}